\documentclass[aip,jcp,showpacs]{revtex4}
\usepackage{bm}
\usepackage{epsfig,amsopn}
\usepackage{graphicx}
\usepackage{amsmath,amssymb}
\usepackage{natbib}
\usepackage{color}
\usepackage{latexsym}
\usepackage{amsfonts}
\usepackage{amssymb}
\usepackage{comment}

\newcommand{\bea}{\begin{eqnarray}}
\newcommand{\eea}{\end{eqnarray}}

\def\beq{\begin{eqnarray}}
\def\eeq{\end{eqnarray}}

\begin{document}
\title{
Heat transfer in the spin-boson model: A comparative study in the incoherent tunneling regime}
%
\author{Dvira Segal}
\affiliation{Chemical Physics Theory Group, Department of Chemistry, University of Toronto,
80 Saint George St. Toronto, Ontario, Canada M5S 3H6}

\date{\today}

\begin{abstract}
We study the transfer of heat in the non-equilibrium spin-boson model
with an Ohmic dissipation. In the non-adiabatic limit we derive a formula
for the thermal conductance based on a rate equation formalism at the level
of the non-interacting blip approximation, valid for temperatures $T>T_K$,
with $T_K$ as the Kondo temperature.
We evaluate this expression analytically assuming either weak or strong couplings, and demonstrate that
our results agree with exact relations.
Far-from-equilibrium sitautions are further examined,
showing a close correspondence to the linear response limit.
\end{abstract}


\maketitle

\section{Introduction}

The spin-boson (SB) model, with a two-level system immersed in a dissipative thermal environment, can
describe different physical problems: electron transfer in condensed phases \cite{weiss},
molecular electronic conduction \cite{nitzan},
the Kondo physics \cite{legget}, and the decoherence behavior of superconducting
qubits \cite{legget,karyn}. An extension of this model, coupling
the spin subsystem to two thermal reservoirs at different temperatures, has been suggested as a minimal
model for exploring the phenomenology of quantum heat transfer in anharmonic junctions \cite{sn05}.
We refer below to this extension as the ``non-equilibrium spin-boson model"  (NESB),
and focus on its heat transfer characteristics in the steady-state limit.
This model complements other descriptions of quantum heat transport in low dimensions \cite{dhar08,wangrev,lirev12},
particularly demonstrating the thermal diode effect \cite{chang06,giazoto}.

The behavior of the NESB model can be explored
by developing open quantum systems methodologies to the non-equilibrium (two-bath) case. Recent studies worked out such
generalizations on the basis of perturbative quantum master equation tools \cite{sn05,segal06,thingna12,redfield,ro11},
the Keldysh Green's function formalism,
\cite{vtw10,neGF,Aviv}, and the non-interacting blip approximation \cite{sn05,ns11}.
Similarly, brute force numerically exact simulations of the SB model
have been advanced to explore the heat transfer dynamics in non-equilibrium settings:
the multilayer multiconfiguration time-dependent Hartree theory
 \cite{kirilMCTDH}, influence
functional path integral techniques \cite{segal13} and Monte-Carlo simulations \cite{saitoMC}.
Other related treatments include
the atomistic Green's function approach \cite{mingo,mingorev}, the
generalized quantum Langevin equation \cite{dharstat,yerkes}, and self-consistent extensions,
incorporating (effective) anharmonicities \cite{dharstat,malay,tulkki}.
These developments are not trivial: The dissipative dynamics of a subsystem
is reached by time-evolving  its (reduced) density matrix. In contrast,
the operator describing the heat current involves degrees of freedom of the subsystem and
 reservoirs. Thus, one should first work out a closed-workable expression for the heat current,
such that it only depends on degrees of freedom of the subsystem.

A formally exact construction for the heat current in quantum junctions has been derived in Ref. \cite{vtw10}
from the perturbation expansion of the non-equilibrium Green's function.
This formula  
expresses the heat current of the NESB model in correlation functions of the
spin subsystem. This expression can be used to obtain the Redfield Born-Markov result \cite{vtw10}, but more
fundamentally, its linear response limit 
was evaluated with Monte-Carlo simulations,
to explore signatures of Kondo physics in thermal conduction \cite{saitoMC}.

In parallel to these developments, in a series of recent papers we
adopted the non-interacting blip approximation (NIBA) \cite{weiss}
and introduced an approximate expression for the heat current, valid in the non-adiabatic limit
and potentially far-from-equilibrium \cite{sn05,segal06,ns11}.
This was achieved in the picture of the polaron-shifted NESB Hamiltonian. We energy-resolved
the quantum master equation in the non-adiabatic limit,
then derived the cumulant generating function of the system \cite{sn05,ns11}.
The resulting expression for the heat current, a convolution-like integral, conjoins
transition rate constants between the two spin states.
This approach thus builds on the analytical and numerical machinery
developed to treat electron transfer reactions at the level of the Fermi Golden Rule  \cite{weiss,nitzan}.
Other advantages of this heat-transfer NIBA formalism are potential extensions
to handle multi-state junctions \cite{segal06} and far-from-equilibrium situations \cite{ns11}.

In this paper we focus on the behavior of the thermal conductance in the NESB
model at high temperatures.
Our goal is to prove that the approximate heat-transfer NIBA treatment \cite{sn05,ns11} provides analytic results
in agreement with exact simulations \cite{saitoMC}, in the right limits.
This correspondence establishes the heat-transfer NIBA formalism, which
could be advanced to treat more complex multi-state junctions.
Furthermore, we apply our method away from linear response
and discuss the behavior of the current in the limits of weak and strong system-bath couplings.
%

The paper is organized as follows. In Sec. \ref{model} we present the NESB model. We further
include the exact formula for the thermal conductance and its high temperature limits by following Ref. \cite{saitoMC}.
In Sec. \ref{NIBA} we present the approximate heat-transfer NIBA expression, use it to derive closed forms for
the thermal conductance at weak and strong couplings,
then include numerical results. Sec. \ref{Summ} summarizes our work.


\section{Model and exact current formula}
\label{model}

The model comprises a two-state system (spin) attached to two bosonic reservoirs ($\nu=L,R$), and
we focus here on the unbiased case with degenerate spin levels,
\beq
H\!=\!{\hbar\Delta\over 2} \sigma_x +
\sum_{\nu,k} \left[
{\hbar\sigma_z\over 2}
\lambda_{k,\nu}
(b_{k,\nu}^{\dagger} + b_{k,\nu}) +\hbar \omega_{k} b_{k,\nu}^{\dagger} b_{k,\nu} \right].
\label{eq:ham}
\eeq
The operators $\sigma_{i} \, (i=x,y,z)$ are the Pauli matrices,
$\Delta$ stands for the tunneling frequency between the spin states and
$b_{k,\nu}^{\dagger}$ ($b_{k,\nu}$) is the creation (annihilation)
operator of a boson (e.g. phonon) with a wave-number $k$ in the $\nu$ reservoir.
The interaction of the subsystem with the environment can be characterized
by the spectral density function
\beq
J_{\nu} (\omega) = \sum_{k} \lambda_{k,\nu}^2 \, \delta (\omega - \omega_{k}).
\eeq
Below we assume that an Ohmic function characterizes both reservoirs,
\beq
J_{\nu} (\omega) = 2\alpha_{\nu} \omega e^{-\omega/\omega_c}.
\label{eq:spect}
\eeq
Here $\alpha_\nu$ is a dimensionless interaction parameter
between the spin subsystem and the $\nu$ reservoir, and we introduce the definition $\alpha\equiv\alpha_L + \alpha_R$.
For simplicity, the cutoff frequency $\omega_c$ is taken identical at both baths.
The two reservoirs are separately prepared in a canonical-equilibrium state of temperature $T_{\nu}$.
At time $t=0$ we couple the two baths  indirectly through the subsystem, then wait for steady-state to set in.

An exact  Meir-Wingreen-like heat current
expression \cite{mw92} has been derived in several works \cite{vtw10,wwz06,ro08,saito}.
In the steady-state limit it can be regarded as a many-body extension of
the Landauer formula \cite{rk98}. Assuming a sharp cutoff at $\omega_c$ [(rather than the exponential form
of Eq. (\ref{eq:spect})], this Meir-Wingreen-type heat current was written in Ref. \cite{saitoMC} as
\beq
j_q \!\!
&=& \!\! {\hbar^2 \alpha_L\alpha_R \over (\alpha_L+\alpha_R)} \!
\int_{0}^{\infty} \!\!\! d \omega \,  \omega
\chi'' (\omega ) \tilde{I} (\omega) \left[ n_L(\omega )- n_R(\omega ) \right]. ~~~
\label{eq:exact}
\eeq
Here $\tilde I(\omega)=\omega\theta(\omega_c-\omega)\theta(\omega)$,
 $\chi'' (\omega)$ is the imaginary part of the Fourier's transform
of the response function of the spin,
$\chi(t,t') = i\hbar^{-1}\theta (t-t') \langle [\sigma_z (t) , \sigma_z (t') ]\rangle \,$, and
$n_{\nu}(\omega)=[e^{\hbar\omega/k_BT_{\nu}}-1]^{-1}$ stands for the Bose-Einstein distribution function.
In the linear response regime
$j_q\sim\kappa(T_L-T_R)$, and we extract the thermal conductance  from the relation
\bea
\kappa = \frac{d j_q}{ d T_L} \Bigg|_{ T_L\to T_R=T}.
\label{eq:kappadef}
\eea
Eq. (\ref{eq:exact}) then reduces to
\beq
\kappa &=& k_B \hbar {\alpha_L\alpha_R  \over(\alpha_L+\alpha_R)}
\int_{0}^{\omega_c} \!\! d \omega
S_{\alpha} (\omega )\, \omega^2
\Bigl[ {\beta\hbar\omega/2  \over \sinh (\beta\hbar\omega /2 ) } \Bigr]^2,
\label{eq:kappa}
\eeq
with $k_BT=\beta^{-1}$ and the spectral function
$S_{\alpha}(\omega)\equiv\chi''(\omega )/\omega$.
Eq. (\ref{eq:kappa}) was used in Ref. \cite{saitoMC}
as the basis for exact numerical simulations:
The spin response function was evaluated by a
Monte-Carlo method,
performed by noting that the equilibrium partition function of the SB model
can be mapped onto the one-dimensional Ising model  with long range interactions \cite{saitoMC}.
These simulations had indicated that the thermal conductance follows the scaling form
\beq
\kappa &=& \frac{4k_B^2T_K}{\hbar} \frac{\alpha_L\alpha_R}{(\alpha_L+\alpha_R)^2} \, f(\alpha, T/T_K ),
\label{eq:scaling}
\eeq
where $f(T/T_K)\propto (T/T_K)^3$ at low temperatures, $T\ll T_K$; $T_K$ is the Kondo temperature in the system,
a function of the microscopic parameters
$\Delta$, $\omega_c$ and $\alpha$.  In the range $0<\alpha<1$
it is given by \cite{legget,weiss,saitoMC}
\beq
T_K =\frac{\hbar \Delta}{k_B}  \left( \frac{\Delta}{\omega_c}\right)^{\alpha/(1-\alpha)}[\Gamma(1-2\alpha)\cos(\pi\alpha)]^{1/2(1-\alpha)}.
\label{eq:kondo}
\eeq
When $\alpha\geq1$, $T_K=0$. Here $\Gamma(x)$ represents the  Gamma function.

We now discuss the high temperature limit, $T>T_K$, of Eq. (\ref{eq:kappa}) by following Ref. \cite{saitoMC}.
In the weak coupling limit we use
the zeroth-order spin correlation function, namely, the isolated spin solution.
This exercise results in the form
\beq
\kappa_{\alpha\ll1} &=& k_B{ \alpha_L\alpha_R  \over (\alpha_L+\alpha_R) }
 { \pi \Delta \over 2 n(\Delta ) +1 }
\Bigl[ {  \beta\hbar\Delta /2  \over \sinh (\beta\hbar\Delta/2  ) } \Bigr]^2 \, ,~~~~
\nonumber\\
&\xrightarrow{\beta\hbar\Delta\ll1}&
\frac{\hbar\pi}{2} \frac{\Delta^2}{T} \frac{\alpha_L\alpha_R}{\alpha_L+\alpha_R}.
\label{eq:wc}
\eeq
Here $n(\omega)=[e^{\beta\hbar\omega}-1]^{-1}$ denotes the Bose-Einstein
distribution function at the inverse temperature $\beta=1/k_BT$.
More generally, 
one can derive a weak-coupling formula for the non-equilibrium
heat current, directly from the Born-Markov quantum master equation \cite{sn05,segal06,ns11},
\bea
j_q=
\hbar\Delta\frac{ \Gamma_L\Gamma_R
[n_L(\Delta)-n_R(\Delta)]}
{\Gamma_L[1+2n_L(\Delta)]+\Gamma_R[1+2n_R(\Delta)]}.
\label{eq:jweak}
\eea
Here $\Gamma_{\nu}(\omega)=\frac{\pi}{2} J_{\nu}(\omega)$ stands for the system-bath interaction frequency,
evaluated at the frequency $\Delta$ in Eq. (\ref{eq:jweak}).

Beyond the weak coupling limit and at high temperatures,
$T_K \ll  T\ll \hbar\omega_c/k_B$,
a closed expression for the thermal conductance is achieved by adopting
the spin spectral function  at the level of the non-interacting blip approximation
 \cite{weiss, legget}
\bea
S_{\alpha} (\omega) \simeq 2\zeta /[(\omega^2 + \zeta^2)\hbar\omega\coth(\beta\hbar\omega/2)],
\eea
where $\zeta \propto (\Delta^2/\omega_c)(\beta\hbar\omega_c)^{1-2\alpha}$; recall that
$\alpha=\alpha_L+\alpha_R$.
Plugging this expression into Eq. (\ref{eq:kappa}) we arrive at the form \cite{saitoMC}
\beq
\kappa &\simeq& {\cal C} \frac{k_B \Delta^2}{ \omega_c}
\left(\frac{k_B T}{\hbar \omega_c}\right)^{2\alpha-1} .
\label{eq:kappa_highT}
\eeq
Here ${\cal C}$ is a prefactor which weakly depends on the coupling strength.
We emphasize that this result was derived from the exact heat current formula (\ref{eq:kappa}),
with an approximate-NIBA spin-spin correlation function.


\section{Approximate heat current formula: Non-interacting blip approximation}
\label{NIBA}

The NIBA scheme is valid in the non-adiabatic limit $\omega_c>>\Delta$. It
can faithfully simulate the SB dynamics at strong system-bath interactions and/or at high
temperatures in the Ohmic case.
It is also exact for the unbiased model at weak damping.
We focus on the occupation of the spin states $p_{1,0}(t)=(1\pm\langle \sigma_z(t)\rangle$)/2,
$\langle\sigma_z(t)\rangle={\rm tr}[\rho(0)\sigma_z(t)]$, $\rho(0)$ is the initial-total density matrix.
Under NIBA they satisfy
 an integro-differential equation 
\cite{weiss}
\bea \frac{dp_1(t)}{dt}&=& -\frac{\Delta^2}{2} \int_{0}^{t}
e^{-Q'(t-\tau)} \cos[ \omega_0(t-\tau)-Q''(t-\tau)] p_1(s) d\tau
\nonumber\\
& +&\frac{\Delta^2}{2} \int_{0}^{t} e^{-Q'(t-\tau)} \cos[
\omega_0(t-\tau)+Q''(t-\tau)] p_0(\tau)d\tau.
\label{eq:P2}
\eea
Here $\omega_0$ stands for the spin spacing in the biased SB model, when augmenting the
Hamiltonian (\ref{eq:ham}) with the term $\hbar\omega_0\sigma_z/2$.
The function $Q(t)=\sum_{\nu}Q_{\nu}(t)$,
$Q_{\nu}(t)=Q_{\nu}'(t)+iQ_{\nu}''(t)$ includes real and imaginary components,
\bea
Q'_{\nu}(t)& = & \int_{0}^{\infty}d\omega\frac{J_{\nu}(\omega)}{\omega^2}[1-\cos(\omega t)] [1+2n_{\nu}(\omega)],
\nonumber\\
Q''_{\nu}(t)& = &  \int_{0}^{\infty}d\omega \frac{J_{\nu}(\omega)}{\omega^2}\sin(\omega t).
\label{eq:Q}
\eea
%
By energy-unraveling equation (\ref{eq:P2}), we
derived in Ref. \cite{ns11} a closed expression for the steady-state heat current, defined
positive when flowing left to right,
\bea j_q =
\left(\frac{\Delta}{2}\right)^2 \frac{\hbar}{2\pi}\int_{-\infty}^{\infty}\omega d\omega
\left[ k_R(\omega)k_L(\omega_0-\omega)p_1^{ss} -
k_R(-\omega)k_L(-\omega_0+\omega)p_0^{ss} \right],
\label{eq:Current} \eea
with $p_{1,0}^{ss}$ as the steady-state population of the spin states.
The elements $k_{\nu}(\omega)$ are related to the single-bath
non-adiabatic (Fermi Golden Rule) transition rate constants, only missing the $\Delta^2$ prefactor,
\bea
k_{\nu}(\omega)=\int_{-\infty}^{\infty}e^{i\omega t} e^{-Q_{\nu}(t)}dt.
\label{eq:rate}
\eea
It can be shown that these terms satisfy the detailed balance relation,
\bea
k_{\nu}(-\omega)=k_{\nu}(\omega)e^{-\beta\hbar\omega}.
\label{eq:ratedb}
\eea
In the unbiased case, $\omega_0=0$, $p_0^{ss}=p_1^{ss}=1/2$, and Eq. (\ref{eq:Current}) reduces to
the compact form
\bea j_q =
\left(\frac{\Delta}{2}\right)^2 \frac{\hbar}{4\pi}\int_{-\infty}^{\infty}\omega
\left[ k_R(\omega)k_L(-\omega) -
k_R(-\omega)k_L(\omega) \right]d\omega.
\label{eq:Currentw0}
\eea
It is interesting to note that the terms $k_{\nu}(\omega)$, which directly correspond to the
transition rates in the {\it biased} SB model, 
serve as the elemental ingredient in the heat current expression, in the {\it unbiased} NESB model.
We can understand this connection by interpreting
Eq. (\ref{eq:Currentw0}) as follows: The total energy current is given by a sum over all possible
energy exchange processes, with the amount of energy $\omega$ transferred from $L$ to $R$; the weight
is given by the combination of terms $k_{\nu}(\pm\omega)$.
Using the linear response definition, Eq. (\ref{eq:kappadef}),
and the detailed balance relationship, we obtain the thermal conductance
%
%
%
\bea
\kappa
&=&
\left(\frac{\Delta}{2}\right)^2 \frac{\hbar}{4\pi}
\int_{-\infty}^{\infty} \omega
\left[ k_R(\omega)\frac{dk_L(-\omega)}{dT_L} -
k_R(-\omega)\frac{d k_L(\omega)}{dT_L} \right]d\omega\Big|_{T_L \to T_R=T}
\nonumber\\
&=&
\left(\frac{\Delta}{2}\right)^2 \frac{\hbar^2}{4\pi k_B T^2}
\int_{-\infty}^{\infty}\omega^2
 k_R(\omega)k_L(-\omega)d\omega.
\label{eq:kappa2}
\eea
In the second line $k_R(\omega)$ and $k_L(\omega)$ are evaluated at the same temperature $T$; these terms may still differ
if $\alpha_L\neq \alpha_R$. Eq. (\ref{eq:kappa2}), a formal expression for the thermal conductance within NIBA,
is the first main result of this work.

The non-adiabatic Golden Rule factors $k_{\nu}(\omega)$ of Eq. (\ref{eq:rate})
have been the focus of many studies, particularly in the context of electron transfer reactions
in solution \cite{weiss,nitzan}.
We can now build on these results, and obtain the heat current in different limits.
For example, in the Ohmic case, in the scaling regime, $k_BT,\hbar\omega < \hbar\omega_c$, it can be shown that
\cite{weiss}
\bea
k_{\nu}(\omega)=
\frac{1}{\omega_c} \left( \frac{\hbar \omega_c}{2\pi k_BT} \right)^{1-2\alpha_{\nu}}
\frac{|\Gamma(\alpha_{\nu}+i\hbar \omega / 2\pi k_BT)|^2}
{\Gamma(2\alpha_{\nu})}e^{\hbar\omega/2k_BT},
\label{eq:k1}
\eea
with the Gamma function $\Gamma(x)$.
This result was derived for the SB model, with $\omega$ serving as the energy gap between the spin states,
limited to small values, $\omega<\omega_c$.
It can be used in Eq. (\ref{eq:kappa2}), replacing both $k_L(-\omega)$ and $k_R(\omega)$.
Since the heat current is dominated by bath modes of significant thermal occupation,
it is sufficient to evaluate the integral up to  the frequency $\sim k_BT/\hbar<\omega_c$ where Eq. (\ref{eq:k1}) holds.
%
At strong coupling and high
temperatures we ignore the $\omega$ dependence within the Gamma function,
and we end up with the relation
\bea
\kappa &\simeq& \mathcal A \hbar^2\left(\frac{\Delta}{\omega_c}\right)^2
\frac{1}{k_BT^2}
\left(\frac{\hbar \omega_c}{ k_B T}\right)^{2-2\alpha_L-2\alpha_R}
\int_0^{k_BT/\hbar} \omega^2d\omega
\nonumber\\
&\simeq& \mathcal A
\frac{k_B\Delta^2}{\omega_c}
\left(\frac{\hbar \omega_c}{ k_B T}\right)^{1-2\alpha}.
\label{eq:kapNiba}
\eea
%
Here  $\mathcal A$  is a constant which weakly depends on $\alpha$
through the Gamma function in Eq. (\ref{eq:k1}).
 This result, our second  contribution,
agrees with  Eq. (\ref{eq:kappa_highT}) and with
exact numerical simulations \cite{saitoMC}.
One should note that the derivation of Eqs. (\ref{eq:kappa_highT}) and (\ref{eq:kapNiba}) differs:
While the former is derived from the exact formula, only
the spin susceptibility is approximated based on the NIBA,
Eq. (\ref{eq:kapNiba}) results from the approximate heat current expression (\ref{eq:Current})
with the non-adiabatic rates $k_{\nu}(\omega)$.
In the weak coupling limit, $\alpha_{\nu}\ll1$,  the Fermi Golden Rule expression
(\ref{eq:rate}) reduces to \cite{weiss}
%
\bea
k_{\nu}(\omega)=
\left(\frac{2\pi k_BT}{\hbar\Delta}\right)^{2\alpha_{\nu}} \frac{4\pi \alpha_{\nu}\omega}{(2\pi \alpha_{\nu}k_BT/\hbar)^2+\omega^2}
\frac{1}{1-e^{\hbar\omega/k_BT}}.
\eea
We use this form to replace both
$k_L(-\omega)$ and $k_R(\omega)$ in Eq. (\ref{eq:kappa2}), to obtain
\bea
\kappa_{\alpha\ll1} &= &\left(\frac{\Delta}{2}\right)^2 \frac{\hbar}{4\pi}
\int_{-\infty}^{\infty}\frac{(4\pi)^2\alpha_L\alpha_R\omega^3d\omega}{\left[(2\pi\alpha_Lk_BT/\hbar)^2+\omega^2\right]\left[(2\pi\alpha_Rk_BT/\hbar)^2+\omega^2\right]} \frac{dn(\omega)}{dT}.
\eea
Recall that $n(\omega)$ is the Bose-Einstein distribution function.
In the high temperature limit, $dn/dT\sim k_B/\hbar\omega$.
We use the definite integral
$\int_{-\infty}^{\infty} \frac{x^2}{(a^2+x^2)(b^2+x^2)}dx = \frac{\pi}{a+b}$,
and reach the thermal conductance
\bea\kappa_{\alpha\ll1} =\frac{\hbar \pi}{2}
\frac{\Delta^2}{T}
\frac{\alpha_L\alpha_R}{(\alpha_L+\alpha_R)},
\label{eq:kapwNiba}
\eea
in agreement with the high temperature limit of Eq. (\ref{eq:wc})
and other weak-coupling schemes \cite{sn05,segal06}.
We have thus confirmed that the NIBA heat current formula (\ref{eq:Current}) produces the
high temperature limit of the thermal conductance, in agreement with exact Monte-Carlo simulations \cite{saitoMC}
and Born-Markov weak coupling expansions \cite{sn05}.
Analytic results in the classical ``Marcus" regime, $k_BT>\hbar \omega_c$,
were discussed in Refs. \cite{sn05,ns11,segalspin14} and we do not repeat them here.

\begin{figure}[htbp]
\vspace{0mm} \hspace{0mm}
{\hbox{\epsfxsize=75mm \epsffile{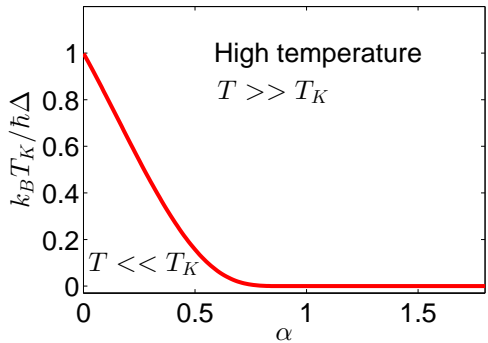}}}
\caption{The Kondo Temperature $T_K$ of Eq. (\ref{eq:kondo}) with
$\alpha_L=\alpha_R$,
$\omega_c=10\Delta$. NIBA results are valid in the high temperature regime.}
\label{Fig0}
\end{figure}

\begin{figure}[htbp]
\vspace{0mm} \hspace{0mm}
{\hbox{\epsfxsize=70mm \epsffile{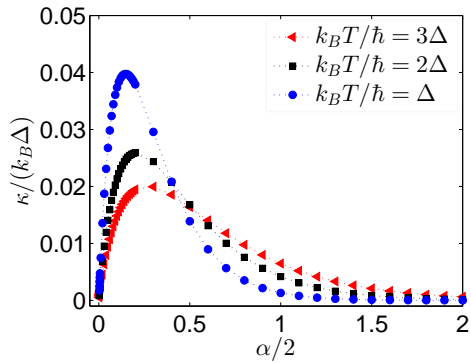}}}
\caption{
Thermal conductance as a function of $\alpha$,
calculated from Eq. (\ref{eq:kappa2}) with
the non-adiabatic rates (\ref{eq:rate}), assuming an Ohmic form and $\alpha_L=\alpha_R$,
$\omega_c=10\Delta$.}
\label{Fig1}
\end{figure}

\begin{figure}[htbp]
\vspace{0mm} \hspace{0mm}
{\hbox{\epsfxsize=80mm \epsffile{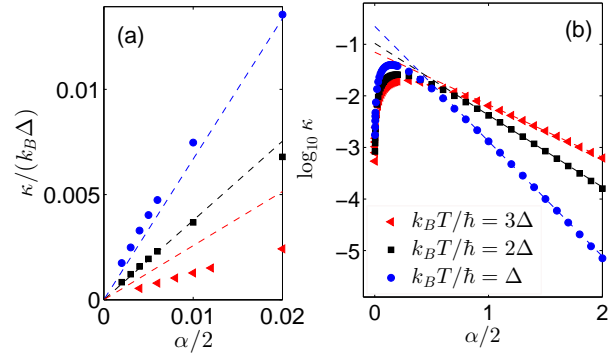}}}
\caption{
Thermal conductance as a function of $\alpha$, with the same data as in
 Fig. \ref{Fig1}.
(a) Weak coupling limit:
 NIBA expression Eq. (\ref{eq:kappa2}) (symbols),
 Born-Markov result (\ref{eq:wc}) (dashed lines).
(b)
Strong coupling limit, demonstrating the scaling (\ref{eq:kapNiba}).
The legend describes both panels.
}
\label{Fig2}
\end{figure}

\begin{figure}[htbp]
\vspace{0mm} \hspace{0mm}
{\hbox{\epsfxsize=65mm \epsffile{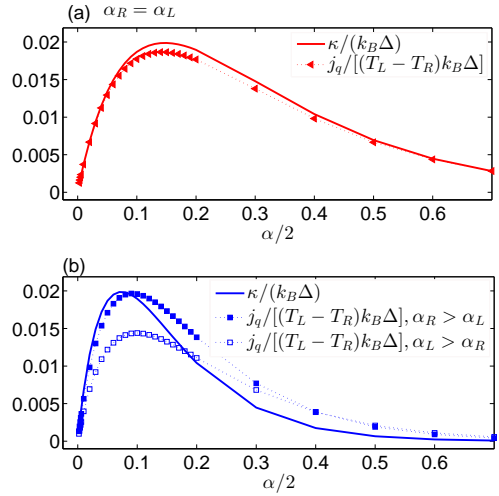}}}
\caption{The NESB model
beyond linear response:
Heat current  (\ref{eq:Current}) divided by the temperature difference
and thermal conductance, Eq. (\ref{eq:kappa2}) at
(a) $\alpha_R=\alpha_L$,
(b) $\alpha_L\neq\alpha_R$. In both cases
$k_BT_L/\hbar\Delta=1.5$, $k_BT_R/\hbar\Delta=0.5$, $\omega_c=10\Delta$.
The thermal conductance is
evaluated at the average temperature $T=(T_L+T_R)/2$.
}
\label{Fig3}
\end{figure}


We display next numerical results of the thermal conductance using NIBA,  
Eq. (\ref{eq:kappa2}) with the rates (\ref{eq:rate}). In Fig. \ref{Fig0} we plot the Kondo temperature
as a function of the coupling parameter $\alpha$. This figure identifies
the high temperature region $T>T_K$ in which NIBA simulations are meaningful.
The thermal conductance is presented in Figs. \ref{Fig1}-\ref{Fig2}, and we confirm 
that the NIBA formula results in correct forms at weak and strong couplings.
Specifically, at strong coupling, the relation
$\log\kappa\propto 2\alpha\log(\hbar\omega_c/k_BT)$ is obeyed;
using $k_BT=1,2,3$ $\hbar\Delta$
and $\omega_c=10\Delta$  
we extract the (large $\alpha$) respective slopes $2.23, 1.40, 1.02$
from Fig. \ref{Fig2}. 
This closely agrees with the theoretical values of $2\log_{10}(\hbar\omega_c/k_BT)=2.00,1.40,1.05$.

In Fig. \ref{Fig3} we explore the behavior of the heat current beyond linear response, adopting Eq. (\ref{eq:Current}).
For spatially symmetric systems deviations from equilibrium manifest themselves predominantly in the crossover
(weak-to-strong) region. When asymmetry in the form $\alpha_L\neq \alpha_R$ is implemented, deviations are more pronounced
since $(T_L-T_R)^{2n}$ terms, $n=1,2,..$,  responsible for thermal rectification, contribute. Particularly,
at weak coupling the junction better conducts when it is coupled weakly to the hot terminal, and more strongly to the cold one.
Formally, we expand the current in powers of $\delta T=T_L-T_R$,
\bea
j_q=\kappa\delta T +  \kappa_2 \frac{\delta T^2}{T}+
\kappa_3 \frac{\delta T^3}{T^2}+....
\eea
Considering the spatially symmetric case at strong coupling, we can readily prove that
the heat current $j_q$ follows the functional form (\ref{eq:kapNiba}), preserving linear response characteristics:
We plug Eq. (\ref{eq:k1}) in the heat current expression (\ref{eq:Currentw0}),
 replacing the four terms $k_L(\pm\omega)$ and $k_R(\pm \omega)$. In the strong coupling limit we ignore the $\alpha$ dependence
of the Gamma function. We temperature-bias the baths in a symmetric manner,
 $T_{L}=T+\delta T/2$ and $T_R=T-\delta T/2$, and reach the relation
\bea j_q &\simeq&
\left(\frac{\Delta}{2\omega_c}\right)^2 \frac{\hbar}{4\pi}
\left[\frac{(\hbar\omega_c)^2}{(2\pi k_B)^2(T^2-\delta T^2/4)}\right]^{1-\alpha}
\nonumber\\
&\times&
\int_{-\infty}^{\infty}\omega \left[e^{\frac{\hbar\omega}{2k_B}(\frac{1}{T_R} - \frac{1}{T_L})}
-e^{-\frac{\hbar\omega}{2k_B}(\frac{1}{T_R} - \frac{1}{T_L})} \right]
d\omega.
\label{eq:Current3}
\eea
We now expand the exponential functions in the integrand in powers of $\delta T$,
\bea
&& \left[e^{\frac{\hbar\omega}{2k_B}\left(\frac{1}{T_R} - \frac{1}{T_L}\right)}
-e^{-\frac{\hbar\omega}{2k_B}\left(\frac{1}{T_R} - \frac{1}{T_L}\right)} \right]
\nonumber\\
&&\sim \frac{\delta T}{k_BT^2}\hbar\omega +\delta T^3 \hbar\omega \frac{6(k_BT)^2+(\hbar\omega)^2}{24k_B^3T^6} + ...,
\eea
and perform the frequency integration with an upper limit $k_BT/\hbar$. We immediately reach a form
parallel to Eq. (\ref{eq:kapNiba}),
\bea
j_q\simeq k_B\frac{\Delta^2}{\omega_c} 
\left(\frac{\hbar\omega_c}{k_BT}\right)^{1-2\alpha}
\left[c_1\delta T  + c_3\frac{\delta T^3}{T^2} +... \right],
\eea
with the numeric factors $c_{1}$ and $c_3$.
This result does not quantify the importance of nonlinear effects, the ratio $c_3/c_1$. It only points out
that high-order conductances maintain the form of the linear response term.
A similar analysis can be performed in the weak coupling limit, to confirm
that Eq. (\ref{eq:kapwNiba}) describes high-order conductances.

The third principal result of this paper is thus that equations (\ref{eq:kapNiba})
 and (\ref{eq:kapwNiba}) portray the behavior of high-order conductances in spatially symmetric systems, at weak and strong coupling, respectively.
It is  significant to note that the thermal diode effect is optimized in a certain region, $0.1<\alpha<0.3$, see Fig. \ref{Fig3}(b).

\section{Summary}
\label{Summ}

We considered the problem of thermal transport in the non-equilibrium spin-boson model and
showed that a NIBA-based formula for the thermal conductance, justified in the non-adiabatic limit ($\Delta<\omega_c$),
provides analytic results in agreement with exact simulations \cite{saitoMC} in the high temperature limit $T>T_K$.
Away from equilibrium, we found that nonlinear effects show up in the weak-intermediate interaction regime,
$\alpha=0.1-0.3$, and that their functional form follows the linear response limit.

We conclude by emphasizing the utility of the NIBA heat current formula (\ref{eq:Current})
in its $T>T_K$ regime of applicability:
(i) It is based on the
Fermi Golden Rule, extensively investigated
in the context of electron transfer reactions \cite{weiss,nitzan}. This allows us to
adopt established expressions for the transition rates, to obtain transport characteristics.
(ii) The NIBA formula can be used with minimal adjustments to handle other thermal baths beyond the harmonic case, e.g.,
spin baths \cite{segalspin14}.
(iii) It is valid beyond linear response, to provide the heat current
in systems far-from-equilibrium.
(iv) Equation (\ref{eq:Current})
can be extended to simulate multi-state junctions \cite{segal06}; in such cases exact simulations are impractical.
Future work will be devoted to time-dependent effects for
addressing quantum heat pumping problems \cite{jie10,jie13,uchi}. It is also of interest
to obtain closed forms for the current cumulants \cite{CGF},
and understand how the current-noise scales with system-bath coupling far from equilibrium.

\noindent\\

\begin{acknowledgments}
Support from an NSERC discovery grant is acknowledged.
\end{acknowledgments}


\end{document}